\documentstyle[psfig]{article}
\newcommand{\be}{\begin{equation}}
\newcommand{\ee}{\end{equation}}
\newcommand{\beqas}{\begin{eqnarray*}}
\newcommand{\eeqas}{\end{eqnarray*}}
\newcommand{\beqar}{\begin{eqnarray}}
\newcommand{\eeqar}{\end{eqnarray}}

\begin{document}
\title{Apparent multifractality\\in financial time series}

\author{Jean-Philippe
Bouchaud$^{\dagger,*}$ Marc Potters$^\dagger$ and Martin Meyer$^\dagger$}
\date{{\small $^\dagger$ Science \& Finance, 109-111 rue Victor Hugo,
92532
Levallois {\sc cedex}, FRANCE;\\ http://www.science-finance.fr\\
$^*$ Service de Physique de l'\'Etat Condens\'e,
 Centre d'\'etudes de Saclay,\\
Orme des Merisiers,
91191 Gif-sur-Yvette {\sc cedex}, FRANCE\\}
\today}
\maketitle

\begin{abstract}
We present a exactly soluble model for financial time series that mimics
the long range volatility correlations known to be
present in financial data. Although our model is `monofractal' by
construction, it shows apparent multiscaling as a result of a slow
crossover phenomenon on finite time scales. Our results suggest that
it might be hard to distinguish apparent and true multifractal behavior
in financial data. Our model also leads to a new family of stable laws
for sums of correlated random variables.
\end{abstract}

Many time series exhibit interesting {\it scaling}\/ properties. This
means  that if $x(t)$ denotes the time series, the probability
distribution of the  variations $\delta_T x=x(t+T)-x(t)$, rescaled by a
lag-dependent factor $\xi(T)$,  can be written as:
\be\label{scaling}
P(\delta_T x,T)= \frac{1}{\xi(T)} {\cal F}\left(\frac{\delta_T
x}{\xi(T)}\right),
\ee

where ${\cal F}(u)$ is a time independent scaling function. For example,
if $x(t)$ is constructed by summing independent identically distributed
random variables with finite variance, one has $\xi(T)=\sigma\sqrt{T}$ and
${\cal F}(u)=\exp(-u^2/2)/\sqrt{2\pi}$ for  large $T$. Note that often the
`time' is actually a space coordinate, as it is the case in the analysis
of turbulent velocity fields (where $x$  is the fluid velocity)
\cite{turb}, or fracture  surfaces (where $x$ is the height of the
profile) \cite{EB}. Equation (\ref{scaling}) implies that all moments of
$\delta_T x$ that are finite, scale similarly:
\be
m_q \equiv \langle |\delta_T x|^q \rangle = A_q \xi(T)^q,
\ee
where $A_q$ is a q-dependent number. Very often, $\xi(T)$  behaves as a
simple power-law: $\xi(T)\propto T^\zeta$. In this case of a {\it
monofractal} process, one therefore has, $m_q \propto T^{\zeta_q}$, with
$\zeta_q \equiv  q\zeta$.

This is however not the only possibility, and in some cases, one can
observe {\it multifractal} scaling, in the sense that  $m_q \propto
T^{\zeta_q}$, with $\zeta_q \neq q\zeta$. Such a possibility has  been
advocated for turbulent velocity fields \cite{Mandel,Frisch,exp,Brachet1}
and, more recently, for financial  time series
\cite{Ghash,Mandel2,Lovejoy,Brachet2}. In the case of turbulence, there
is strong theoretical evidence in favor of such a multifractal behavior
\cite{turb,Frisch}.  One can actually analytically derive a non trivial
function $\zeta_q$ within a simple  (`passive scalar') model, which is
thought to retain some essential features of real  turbulence
\cite{Scalar}. The situation is much less clear in the case of financial
markets, where  the only evidence is based on the empirical analysis of
the moments of several time series (typically currencies or stock
indices). The idea of {\it multiplicative cascades},  which is at the
heart of the arguments in favor of multiscaling in  turbulence, is not
easily applicable to price time series (see, however,  \cite{Arneodo}).

In this note, we study to which extent empirical studies on multiscaling
behavior in finance are sensitive to crossover behavior that results in
{\it apparent} multiscaling, even though the studied process is a
monofractal. To that end we present a soluble model that is based on the
study  of financial time series. In the model, the `volatility' (or the
variance) of the elementary price increments is a random variable with
long range correlations, which have been shown to be present in financial
time series \cite{Arneodo,Dacorogna93,Ding93,CPB,Stanley}. The
model is an
exact monofractal, but nevertheless it leads to an apparent multiscaling
behavior \cite{TFR,JPMP}. As we argue below, one finds effective
exponents  $\tilde\zeta_q \neq \zeta_q$ due to a very long crossover
effect, which leads to a systematic negative correction to the true
asymptotic behavior $\zeta_q = q\zeta$. The correction grows with $q$ and
results thereby in a nontrivial functional form of $\tilde\zeta_q$. The
numerical simulation of such a model, which mimics quite well the observed
behavior  of real prices, accurately reproduces the published data in
favor of multiscaling  in financial markets.

The model that we propose is also interesting in its own right. As a
function of the strength of the correlations, we find a transition between
a simple Gaussian behavior for the scaling function ${\cal F}$ (together
with the usual $\sqrt{T}$ scaling for $\xi(T)$) for weak correlations, to
a {\it new}\/ family of stable laws (with a non trivial scaling of
$\xi(T)$) for strong enough correlations. This adds to the very few cases
where the limit distribution for sums of correlated random variables is
exactly known.

Our model is the following: we consider that the time series is built by
summing random
variables:
\be\label{sum}
x(t) = \sum_{k=1}^{N} \delta x_k \qquad N=\frac{t}{\tau},
\ee
where $\tau$ is a microscopic time scale. The elementary increments
$\delta x_k$ are assumed to be given by the product of two independent
random variables, a `sign' $\epsilon_k$ and an amplitude $\sigma_k$:
$\delta x_k=|\sigma_k| \epsilon_k$. The $\epsilon_k$ are furthermore
assumed to be independent Gaussian random variables of variance unity.
$\sigma_k^2$ therefore is the (random) variance of the elementary
increments. We choose the $\sigma_k$'s to be Gaussian random variables of
zero mean \cite{Rq1}, with a correlation function given by $\langle
\sigma_k \sigma_{k+\ell} \rangle=C(|\ell|)$. All moments $m_q$ of $\delta
x_k$ are therefore finite, and the even ones given by
$m_q=[(q-1)!!]^2 C^{q/2}(0)$.

The correlation function $C(|\ell|)$ will be chosen to be a power law for
large arguments: $C(|\ell|) \simeq \gamma |\ell|^{-\nu}$. From several
studies of financial markets, one knows that the variance of the price
increments is indeed also random, with a very slowly decaying time
correlation function:
\be
\langle \sigma_i^2 \sigma_{i+\ell}^2 \rangle -\langle \sigma_i^2
\rangle^2 = 2C^2(|\ell|)
\simeq_{\ell \to \infty} \frac{2 \gamma^2}{|\ell|^{2\nu}},
\ee
where the exponent $\nu$ is found to be on the order of $\nu=0.1$ -- $0.3$
for different markets
\cite{Dacorogna93,Ding93,CPB,Stanley,Arneodo}. The important
point here is that $\nu < 1/2$.

More precisely, we will use the following explicit representation of the
$\sigma$'s:
\be\label{fourier}
\sigma_k = \frac{s_0}{\sqrt{N}} \sum_{m=1}^{N/2} \left(\frac{2 \pi
m}{N}\right)^{\frac{\nu-1}{2}}
\left(z_m e^{\frac{2i\pi mk}{N}}+z_m^* e^{-\frac{2i\pi mk}{N}}\right).
\ee
where the $z_m$'s are independent complex Gaussian variables of unit
variance. In the large $N$ limit, the
resulting correlation function $C(\ell)$ is well defined and decays,
for large $\ell$, as a power-law with $\gamma=s_0^2 \Gamma(\nu) \cos(\pi
\nu/2)/\pi$ while $C(0)$ tends to $s_0^2 \pi^{\nu-1}/\nu$ \cite{finiteN}.

We now turn to the calculation of the cumulants $c_q$ of $P_N(x)$ of
$x(t)$, as given by Eq.\ (\ref{sum}). We will show that these cumulants
scale anomalously with $N$ as soon as $n \geq 2$ (for $0 < \nu < 1/2$).

After making a gauge transform $\epsilon_i \to \mbox{sign}(\sigma_i)
\epsilon_i$, one finds \cite{Rq2}:
\beqar
\lefteqn{ e^{-G_N(z)}\equiv\int dx P_N(x) e^{-izx}}\\\nonumber
&&= \int \prod_{j=1}^N
\left(\frac{d\epsilon_j d\sigma_j}{2\pi}\right) \frac{1}{\sqrt{\det{\bf
C}}}
\exp\left[iz\sum_{j=1}^N \sigma_j \epsilon_j - \sum_{j=1}^{N}
\frac{\epsilon_j^2}{2}
-\sum_{j,k=1}^{N} \frac{\sigma_j ({\bf C}^{-1})_{jk}\sigma_k}{2} \right]
\eeqar
The Gaussian integrals can be easily performed, and leads to the following
expression for the
characteristic function $G(z)$:
\be \label{G}
G_N(z)=\frac{1}{2} \mbox{Tr} \log({\bf{1}} + z^2 {\bf {C}}) =
\sum_{m=1}^{N/2} \log(1+z^2
\tilde C(m)),
\ee
where the bold characters is used for matrices, and where $\tilde C(m)$
are the
eigenvalues of the matrix $\bf C$. From the very construction of the
$\sigma$'s,
one finds that $\tilde C(m)=s_0^2 (2\pi m/N)^{\nu-1}$, each of which is
twofold degenerate.

Expanding $G_N(z)$ in powers of $z$ leads to the cumulants $c_q(N)$ of
$P_N(x)$. All odd order cumulants are zero, while even order cumulants are
given by:
\be
c_q(N)=  2(q-1)! \sum_{m=1}^{N/2} \tilde C(m)^{q/2}.
\ee
Let us first analyze the case $0< \nu <1/2$. In the large $N$ limit, the
sum over $m$ is convergent when $q \geq 4$ and leads to cumulants which do
not  scale as $N$:
\be
c_q \simeq \left(\frac{N}{2\pi}\right)^{{q\over2}(1-\nu)} s_0^q
(q-1)! \sum_{m=1}^{\infty} m^{{q\over2}(\nu-1)},
\ee
while for $q=2$ one finds exactly $c_2=\mbox{Tr}{\bf {C}} = NC(0)$. The
normalized cumulants $c_q/c_2^{q/2}$ therefore behave as $N^{-q\nu/2}$ for
$q \geq 4$ and vanish for $N \to \infty$. This means that the distribution
of $x/\sqrt{N}$ indeed tends to a Gaussian for large $N$. However, the
approach to the Gaussian is slower than for sums of   independent random
variables. In particular, the kurtosis $c_4/c_2^2$ of $P_N(x)$  decays
anomalously, as $N^{-2\nu}$, for $0 < \nu < 1/2$. For larger values of
$\nu$ (i.e.\ when the volatility correlations are weaker), one recovers the
usual scaling of the kurtosis as $1/N$ that  holds for independent
increments. Such an anomalous decay of the kurtosis with time was first
reported for financial time series in \cite{JPMP,CPB}.

The important outcome of the above calculation is that the {\it moments}
$m_q$ of the distribution $P_N(x)$ are not simple power-laws, but sums of
power laws with similar exponents. For example:
\be
m_4 = A_{4,0} N^{2-2\nu} + A_{2,2} N^2,
\label{eq_cross4}
\ee
\be
m_6=A_{6,0} N^{3-3\nu} + A_{4,2} N^{3-2\nu} + A_{2,2,2} N^{3},
\label{eq_cross6}
\ee
where the $A$'s are some coefficients. If $\nu$ is small, these sums of
power-laws can be fitted on many decades with an effective exponent
$\tilde\zeta_q$ such that $m_q \propto N^{\tilde\zeta_q}$. The exponent
$\tilde\zeta_q$ is less than $q/2$, and more and more so as $q$ increases.
However, the true asymptotic behavior predicted by our model is
$\zeta_q=q/2$. In fact, Eqs.~(\ref{eq_cross4}) and (\ref{eq_cross6}) show
that our model (and possibly also real financial data) is better
characterized by the cumulants than by the moments.

In order to illustrate this point numerically, we have generated a
surrogate time series in a way closely related to the above model. Instead
of writing $\delta x_i = |\sigma_i| \epsilon_i$, we have chosen to take
$\delta x_i = \exp(\sigma_i) \epsilon_i$. This leads to a more realistic
time series as compared with real data from financial markets, without
changing the crucial feature of the above model, i.e.\ the very slow decay
of the volatility correlations. In particular, the distribution of the
volatility has a positively skewed, log-normal shape.  The length of our
surrogate time series was taken to be comparable to those analyzed
previously. The moments $m_q=\langle |\delta_T x|^q \rangle$ are plotted
as a function of $T$ for different $q's$, for the choice $\nu=0.2$ (see
Fig.\ 1). The interval of $T$ was chosen to be $\tau=1 \leq T \leq 6000$,
again comparable to the region investigated in previous studies
\cite{Ghash,Mandel2,Lovejoy,Brachet2}. The power law fits are extremely
good, and lead to a function $\tilde\zeta_q$ bending downwards as $q$
increases, shown in Fig.\ 2. For our choice of parameters, the numerical
values of $\tilde\zeta_q$ actually match precisely those reported in
\cite{Brachet2}. We have also checked that the same model, but without
volatility correlations, leads very precisely to
$\tilde\zeta_q=\zeta_q=q/2$.

\begin{figure}
\centerline{\psfig{figure=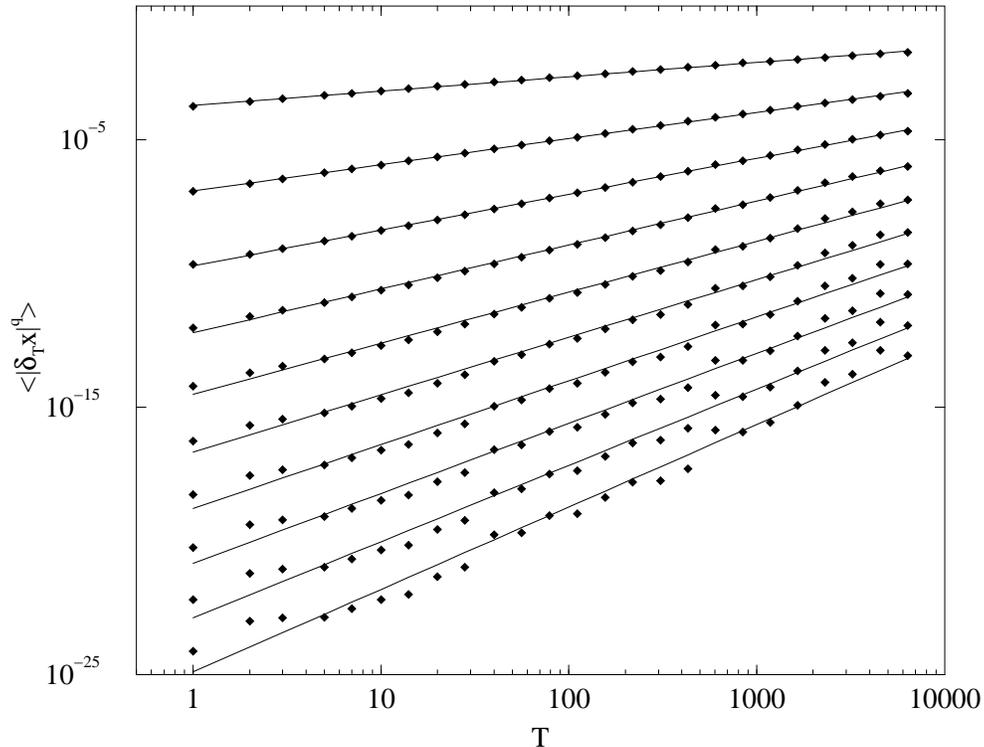,width=13cm}}
\caption{Simulation of our surrogate financial time series and
determination of the scaling of the moments
as a function of window size $T$. From top to bottom $\langle |\delta_T
x|^q \rangle$ for
$q$=1 -- 10. Each moment scales relatively well with $T$ as indicated by
the 10 linear
(log-log) fit lines.
In this simulation the log-volatility follows a correlation Gaussian
process as defined in the text
with $\nu=0.2$. The variance of this process has been set so that the
kurtosis at the unit
time scale be $\kappa_0=65$. This is a typical value for high frequency
financial
data. The simulated data
set contained $500\, 000$ points.}
\label{fig1}
\end{figure}
\begin{figure}
\centerline{\psfig{figure=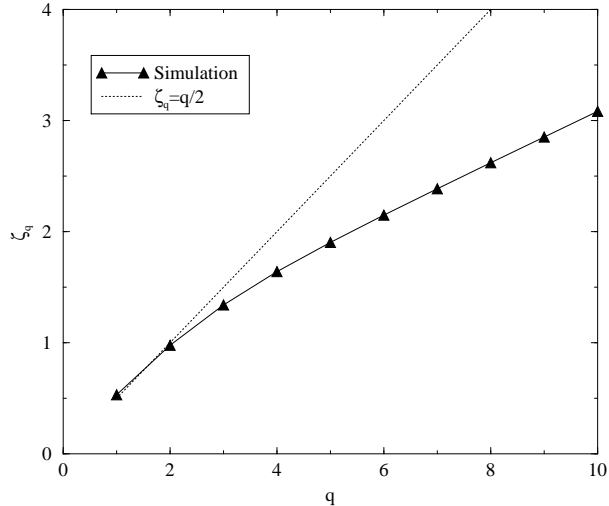,width=8cm}}
\caption{Slope $\zeta_q$ of the fitted data of Fig.\ 1. The deviations
from the
true asymptotic scaling $\zeta_q=q/2$ are quite clear. The value of
$\zeta_q$
precisely match those reported in \protect\cite{Brachet2}.}
\label{fig2}
\end{figure}

We now turn to the case $\nu < 0$. This corresponds to a non stationary
process
for the volatility, which typically {\it grows} with $\ell$ as
$\ell^{-\nu/2}$.
More precisely, from Eq.\ (\ref{fourier}), one can show that $\langle
(\sigma_{k}-\sigma_{k+\ell})^2 \rangle
\propto \ell^{-\nu}$. In this case, after changing variables
to $x=\hat x \sqrt{N^{1-\nu}}$ and $z=\hat z \sqrt{(N/2\pi)^{\nu-1}}$,
one finds that the asymptotic distribution of $\hat x$ has a
characteristic function given by:
\be\label{hatG}
G_\nu(\hat z) = \sum_{m=1}^{\infty} \log\left(1+\hat
z^2 m^{\nu-1} \right)
\ee
(we have set $s_0=1$, which amounts to a change of scale in $x$.).

The above result means that after rescaling by a factor
$\sqrt{N^{1-\nu}}$, the sum of (strongly) correlated random variables
converges to a non-Gaussian distribution ${\cal F}_\nu(\hat x)$, obtained
as the Fourier transform of the exponential of $-G_\nu(\hat z)$ given by
Eq.\ (\ref{hatG}). Since the expansion of $G_\nu(\hat z)$ is regular for
$\hat z \to 0$, all the moments of ${\cal F}_\nu$ are finite.  From the
leading singularity of $G_\nu(\hat{z})$ around $\hat z = \pm i$, one
obtains the asymptotic behavior of ${\cal F}_\nu(\hat x)$ for large
arguments as:
\be {\cal F}_\nu(\hat x) \simeq_{\hat x \to \infty}
e^{-|\hat x|}.
\ee
In the special case where $\nu=-1$, corresponding to a `volatility random
walk', the sum in (\ref{hatG}) can be explicitly performed, and leads to:
\be
G_{-1}(\hat z)=  \log\left(\frac{\sinh(\pi \hat z)}{\pi \hat
z}\right).
\ee
This distribution is shown in Fig.\ 3, together with the predicted
asymptotic behavior (dotted line). The kurtosis of this distribution is
equal to $6/5$.  Interestingly, ${\cal F}_\nu$ has a shape similar to
hyperbolic distributions \cite{Hyperbol} with  exponential tails which
have been proposed in a financial context (see \cite{JPMP}). The
appearance of such laws might thus be related to the existence of
long-ranged correlations in the volatility.

In summary, the purpose of this paper was to show, on an exactly soluble
`stochastic volatility' model, that an apparent multiscaling behavior can
appear as a result of very long transient effects, induced by the long
range nature of the volatility correlations. This model is inspired by
real price time series, and leads to an effective exponent spectrum
$\tilde\zeta_q$ in close correspondence that reported in recent papers on
the subject.
We therefore suspect that indications of mutlifractal behavior found
in financial data might be misleading, as they could be caused by
crossover
effects that do not correspond to the true asymptotic behavior. To check
more carefully for crossover effects, it might be helpful to analyze not
only the moments but also the cumulants in empirical studies.
We also have, {\it en passant}, found a new family of
stable laws for sums of {\it correlated}\/ random variables in the case
where the volatility correlation is growing with time.  It would be very
interesting to characterize the attraction basin of these new stable laws.

\begin{figure}
\centerline{\psfig{figure=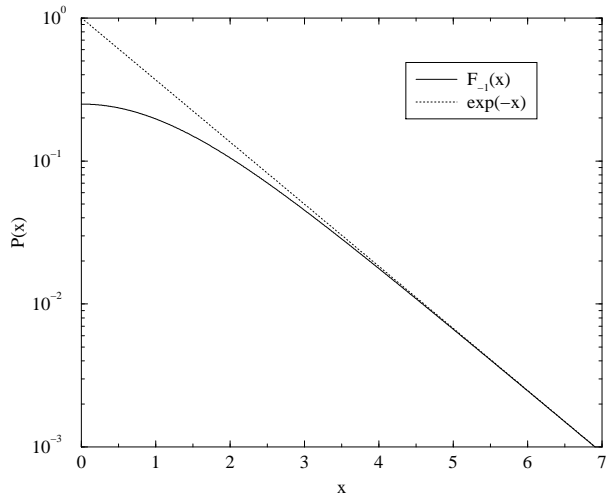,width=8cm}}
\caption{Graph of the stable distribution $P(x)={\cal F}_{-1}(x)$. The
asymptotic behavior
${\cal F}_{-1}(x)\sim \exp(-|x|)$ is shown as the dotted
curve.}
\label{fig3}
\end{figure}

\subsubsection*{Acknowledgements}
We thank M. E. Brachet, P. Cizeau, L. Laloux and A.
Matacz for enlightening discussions.


\begin{thebibliography}{99}
\bibitem{Rq1} One could also choose a non zero mean value for $\sigma$ and
not
take the absolute value. This would still lead to an soluble model.
\bibitem{finiteN} The (time-domain) correlation function defined by
this construction explicitly depends on the number of variables
in the sum $N$. This is not a problem when $\nu>0$, since when $N$ is
large, this correlation function is equal to a well-defined
$N$-independent limit plus
sub-leading correction which can be safely neglected. Therefore, any
subpart
of the series of size $n$ such that $1 \ll n \ll N$ will behave, as a
function
of $n$, as the whole sum.
\bibitem{Rq2} We could have considered the case where the $\epsilon_i$ can
take two values $\pm 1$. The model considered would then be tantamount to
the Ising model in one dimension with long-ranged correlations.
\end{thebibliography}
\end{document}